\input epsf
\catcode`@=11
 \font\tenrm=cmr8 
  \font\sevenrm=cmr7
  \font\fiverm=cmr5
 \font\teni=cmmi8 
  \font\seveni=cmmi7
  \font\fivei=cmmi5
 \font\tensy=cmsy8 
  \font\sevensy=cmsy7
  \font\fivesy=cmsy5
 \font\tenex=cmex8 
 \font\tenbf=cmbx8 
  \font\sevenbf=cmbx7
  \font\fivebf=cmbx5
 \font\tensl=cmsl8 
 \font\tenit=cmti8 
 \skewchar\teni='177 \skewchar\seveni='177 \skewchar\fivei='177
 \skewchar\tensy='60 \skewchar\sevensy='60 \skewchar\fivesy='60
 \textfont0=\tenrm \scriptfont0=\sevenrm \scriptscriptfont0=\fiverm
  \def\rm{\fam\z@\tenrm}
 \textfont1=\teni \scriptfont1=\seveni \scriptscriptfont1=\fivei
  \def\mit{\fam\@ne} \def\oldstyle{\fam\@ne\teni}
 \textfont2=\tensy \scriptfont2=\sevensy \scriptscriptfont2=\fivesy
  \def\cal{\fam\tw@}
 \textfont3=\tenex \scriptfont3=\tenex \scriptscriptfont3=\tenex
 \newfam\itfam \def\it{\fam\itfam\tenit} 
  \textfont\itfam=\tenit
 \newfam\slfam  
  \textfont\slfam=\tensl
 \newfam\bffam \def\bf{\fam\bffam\tenbf} 
  \textfont\bffam=\tenbf \scriptfont\bffam=\sevenbf
  \scriptscriptfont\bffam=\fivebf
 \newfam\ttfam  
 \normalbaselineskip=12pt
 \setbox\strutbox=\hbox{\vrule height8.5pt depth3.5pt width 0pt}
 \normalbaselineskip\rm
\catcode`@=12

\thinmuskip=2mu
\medmuskip=3mu plus 1mu minus 3mu
\thickmuskip=4mu plus 4mu

\newdimen\fullhsize
\fullhsize=18truecm
\def\fullline{\hbox to\fullhsize}
\def\makeheadline{\vbox to 0pt{\vskip-22.5pt
 \fullline{\vbox to8.5pt{}{\pagefont\the\pageno}\runninghead\hfil}\vss}
 \vskip-5truept\hrule\vskip12truept\nointerlineskip}
\def\makefootline{\baselineskip=12pt\fullline{\the\footline}}
\footline={}

\voffset=-0.8truecm
\vsize=24truecm
\hoffset=-1.1truecm
\hsize=18truecm %
\parindent=10truept
\parskip=0pt
\baselineskip=10truept

%
%
\tolerance=10000
\newdimen\colwidth \newdimen\bigcolheight 
\newdimen\pagewidth \newdimen\pageheight 
%
%
\colwidth=\hsize
  \advance\colwidth by -.35truein
  \divide\colwidth by 2
\bigcolheight=\vsize
  \advance\bigcolheight by \vsize
\newdimen\savevsizea \savevsizea=\vsize \advance\savevsizea by 24pt
\newdimen\savevsize \savevsize=\vsize
\newdimen\savehsize \savehsize=\hsize
\def\makefootline{\baselineskip=24pt\hbox to \savehsize{\the\footline}}
\font\sevenrm=cmr7 at 7truept
\font\fiverm=cmr5 at 5truept
\font\seveni=cmmi7 at 7truept
\font\fivei=cmmi5 at 5truept
\font\sevensy=cmsy7 at 7truept
\font\fivesy=cmsy5 at 5truept
\def\Footstrut{\hbox{\vrule height6.72pt depth1.92pt width0pt}}
\def\sevenpoint{\def\rm{\fam0\sevenrm}
	\textfont0=\sevenrm \scriptfont0=\fiverm
	\textfont1=\seveni \scriptfont1=\fivei
	\textfont2=\sevensy \scriptfont2=\fivesy
	\textfont3=\tenex \scriptfont3=\tenex
	\normalbaselineskip=8.64truept
	\normalbaselines\rm}
\def\footnote#1{\edef\@sf{\spacefactor\the\spacefactor}#1\@sf
  \insert\footins\bgroup\sevenpoint
  \interlinepenalty=\interfootnotelinepenalty
  \let\par=\endgraf
  \splittopskip=\ht\strutbox 
  \splitmaxdepth=\dp\strutbox \floatingpenalty=20000
  \leftskip=0pt \rightskip=\colwidth \advance\rightskip by .35in 
	\spaceskip=0pt \xspaceskip=0pt \parindent=1em
  \indent \bgroup\Footstrut #1\aftergroup\Footstrut\egroup
	\let\next}
\pagewidth=\hsize \pageheight=\vsize
\def\onepageout#1{\shipout\vbox{
    \offinterlineskip
    \makeheadline
    \vbox to\savevsizea{#1
	\boxmaxdepth=\maxdepth}
    \makefootline}
    \advancepageno}
  \output{\onepageout{\unvbox255}}
\newbox\partialpage
\def\begindoublecolumns{\begingroup
  \output={\global\setbox\partialpage=\vbox{\unvbox255}}\eject
  \output={\doublecolumnout} \hsize=\colwidth \vsize=\bigcolheight
  \ifvoid\footins\else\advance\vsize by -\ht\footins\fi
  \advance\vsize by -2\ht\partialpage}

\def\doublecolumnout{\dimen0=\pageheight
  \advance\dimen0 by-\ht\partialpage \splittopskip=\topskip
  \ifvoid\footins\setbox0=\vsplit255 to\dimen0\else
   \dimen1=\dimen0
   \advance\dimen1 by-\ht\footins
   \advance\dimen1 by-12pt
   \setbox0=\vbox to \dimen0{\vss\vsplit255 to\dimen1 
	\vskip\skip\footins \kern-3pt \unvbox\footins}\fi
  \setbox2=\vsplit255 to\dimen0
  \onepageout\pagesofar
  \global\vsize=\bigcolheight
  \unvbox255 \penalty\outputpenalty}
\def\pagesofar{\unvbox\partialpage
   \wd0=\hsize \wd2=\hsize \hbox to\pagewidth{\box0\hfil\box2}}
\def\Makevrule{\gdef\pagesofar{\unvbox\partialpage
  \wd0=\hsize \wd2=\hsize \hbox to\pagewidth{\box0\hfil\vrule\hfil\box2}}}
\def\balancecolumns{\setbox0=\vbox{\unvbox255} \dimen0=\ht0
  \advance\dimen0 by\topskip \advance\dimen0 by-\baselineskip
  \divide\dimen0 by2 \splittopskip=\topskip
  {\vbadness=10000 \loop \global\setbox3=\copy0
    \global\setbox1=\vsplit3 to\dimen0
    \ifdim\ht3>\dimen0 \global\advance\dimen0 by1truept \repeat}
  \setbox0=\vbox to\dimen0{\unvbox1}
  \setbox2=\vbox to\dimen0{\unvbox3}
  \global\output={\balancingerror}
  \pagesofar}
\newhelp\balerrhelp{Please change the page
                        into one that works.}
\def\balancingerror{\errhelp=\balerrhelp
        \errmessage{Page can't be balanced}
        \onepageout{\unvbox255}}
%

\font\titlefont=cmbx10 at 12truept
\font\sectionfont=cmbx10
\font\subsfont=cmsl8
\font\headfont=cmbxti10
\font\pagefont=cmbx10

\newcount\refcount \refcount=0
\def\cite#1{\global\advance\refcount by 1\relax {\rm [\the\refcount]}}
\def\nocite#1{\global\advance\refcount by 1\relax}
\newcount\sectionnumber \global\sectionnumber=0
\newcount\subsectionnumber \global\subsectionnumber=0
\def\subsection#1{\global\advance\subsectionnumber by 1\relax
 {\smallskip\noindent{\bf 20.\the\sectionnumber.\the\subsectionnumber.}\quad
 {\subsfont #1}}: }
\def\section#1{\global\advance\sectionnumber by 1\relax
  \global\subsectionnumber=0\relax
  {\smallskip\noindent{\sectionfont 20.\the\sectionnumber.\quad #1}}
  \smallskip\par}
\newcount\fignumber \global\fignumber=0
\def\beginfig{\global\advance\fignumber by 1 \relax
  \begingroup\leftskip=10truept\rightskip=10truept}
\def\endfig{\par\endgroup}
\def\figcap#1{{\beginfig
  \noindent{\bf Figure 20.\the\fignumber:}\quad #1\endfig}}

\def\heading#1{{\centerline{\titlefont#1}}}
\def\WhoDidIt#1{{\noindent#1}}
\def\spose#1{\hbox to 0pt{#1\hss}}
\def\simlt{\mathrel{\spose{\lower 3pt\hbox{$\mathchar"218$}}
     \raise 2.0pt\hbox{$\mathchar"13C$}}}
\def\simgt{\mathrel{\spose{\lower 3pt\hbox{$\mathchar"218$}}
     \raise 2.0pt\hbox{$\mathchar"13E$}}}
\def\lsim{\simlt}
\def\gsim{\simgt}
\def\frac#1#2{{{#1}\over {#2}}}
\def\comma{\, ,}
\def\period{\, .}
\def\etal{{\it et al.}}
\def\ie{{\it i.e.}}
\def\eg{{\it e.g.}}

\def\aj#1,{Astrophys.\ J. {\bf #1},\ }
\def\araa#1,{Ann.\ Rev.\ Astron. Astrophys.\ {\bf #1},\ }
\def\aap#1,{Astron.\ \& Astrophys.\ {\bf #1},\ }
\def\ppnp#1,{Prog.\ in Part.\ Nucl.\ Phys.\ {\bf #1},\ }
\def\prl#1,{Phys.\ Rev.\ Lett.\ {\bf #1},\ }
\def\sci#1,{Science {\bf #1},\ }
\def\nat#1,{Nature {\bf #1},\ }
\def\pp{\par\hangindent=.125truein \hangafter=1\relax}
\def\ppextra{\par\hangindent=.125truein \hangafter=0\relax}
 
\hyphenation{astro-ph}
\hyphenation{an-iso-tro-py}
\hyphenation{an-iso-tro-pies}
\hyphenation{quadru-pole}
\hyphenation{temp-era-ture}
\hyphenation{fluc-tua-tions}
%
%
%

\baselineskip=10truept

%
\def\runninghead{{\headfont\quad 20.\ Cosmic background radiation}}
\overfullrule=5pt


\null
\heading{20.\ COSMIC BACKGROUND RADIATION}
\smallskip

\begindoublecolumns
\null
\vskip -1truecm

\WhoDidIt{Revised October 1997 by G.F. Smoot and D. Scott}

\section{Introduction}
The observed cosmic microwave background (CMB) radiation provides 
strong evidence for the hot big bang.  The success of primordial
nucleosynthesis calculations 
(see Sec.~16, ``Big-bang nucleosynthesis'')
requires a cosmic background radiation (CBR) characterized 
by a temperature $kT \sim 1\,$MeV at a redshift of $z \simeq 10^9$. 
In their pioneering work, Gamow, Alpher, and Herman \cite{Gamow48} realized 
this and predicted the existence of a faint residual relic, 
primordial radiation, with a present temperature of a few degrees.
The observed CMB is interpreted as the current manifestation 
of the hypothesized CBR.

The CMB was serendipitously discovered by Penzias and Wilson \cite{Penzias65}
in 1965.
Its spectrum is well characterized by a $2.73 \pm 0.01\,$K black-body
(Planckian) spectrum over more than three decades in frequency
(see Figs.~20.1 and 20.2).  A non-interacting
Planckian distribution of temperature $T_i$ at redshift $z_i$ transforms 
with the universal expansion
to another Planckian distribution at redshift $z_r$ with temperature 
$T_r / (1+ z_r)= T_i / (1+ z_i)$.
Hence thermal equilibrium, once established (e.g.~at the nucleosynthesis
epoch), is preserved by the expansion, in spite of the fact that photons
decoupled from matter at early times.
Because there are about $10^9$ photons per nucleon,
the transition from the ionized primordial plasma to neutral atoms at
$z\sim1000$ does not
significantly alter the CBR spectrum \cite{peebles93}.

\vskip0.5truecm
\centerline{\epsfxsize=9.0truecm \epsfbox{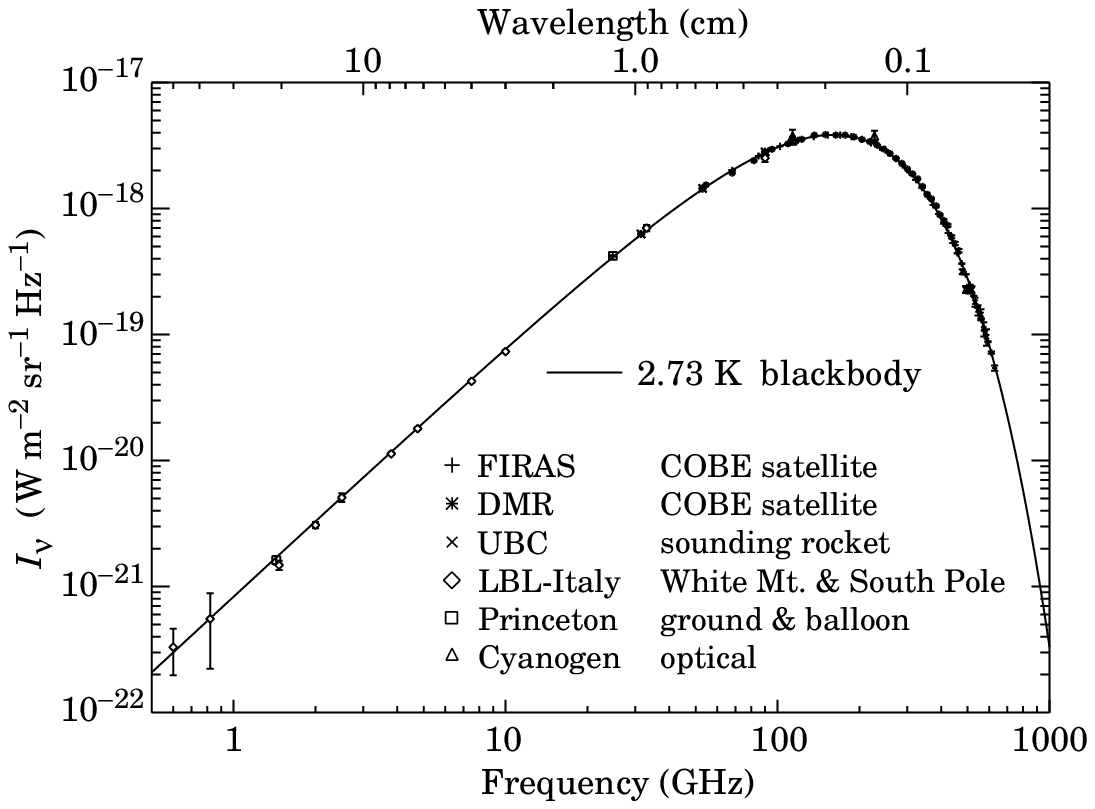}}
\noindent

\figcap{Precise measurements of the CMB spectrum.
The line represents a 2.73~K blackbody,
which describes the spectrum very well,
especially around the peak of intensity.  The spectrum is less well constrained 
at $10\,$cm and longer wavelengths.
(References for this figure are at the end of this section under
``CMB Spectrum References.'')}

\section{The CMB frequency spectrum}
The remarkable precision with which the CMB spectrum is fitted by a Planckian
distribution
provides limits on possible energy releases in the early Universe,
at roughly the fractional level of $10^{-4}$ of the CBR energy,
for redshifts $\lsim10^7$ (corresponding to epochs $\gsim1\,$year).
The following three important classes of spectral distortions (see
Fig.~20.3) generally
correspond to energy releases at different epochs.
The distortion results from the CBR photon
interactions with a hot electron gas 
at temperature $T_e$.


\subsection{Compton distortion} Late energy release ($z\lsim10^5$).
Compton scattering
($\gamma e \rightarrow \gamma' e'$)
of the CBR photons by a hot electron gas creates spectral distortions
by transfering energy from the electrons to the photons.
Compton scattering cannot achieve thermal equilibrium
for $y < 1$, where
$$
y = \int^z_0 ~\frac{ kT_e(z') 
- kT_{\gamma}(z') }{ m_e c^2} \; \sigma_T\; n_e(z') 
\; c\; \frac{dt}{dz'} \;dz'\comma
 \eqno(20.1)
$$
is the integral of the number of interactions, $\sigma_T\; n_e(z)\; c\; dt$, 
times the mean-fractional photon-energy change per collision \cite{sunzel80}.
For $T_e\gg T_\gamma$ $y$ is also proportional to the integral of the
electron pressure $n_e k T_e$ along the line of sight. 
For standard thermal histories
$y < 1$ for epochs later than $z\simeq10^5$.

\vskip-2.1truecm
\centerline{\epsfxsize=10.0truecm \epsfbox{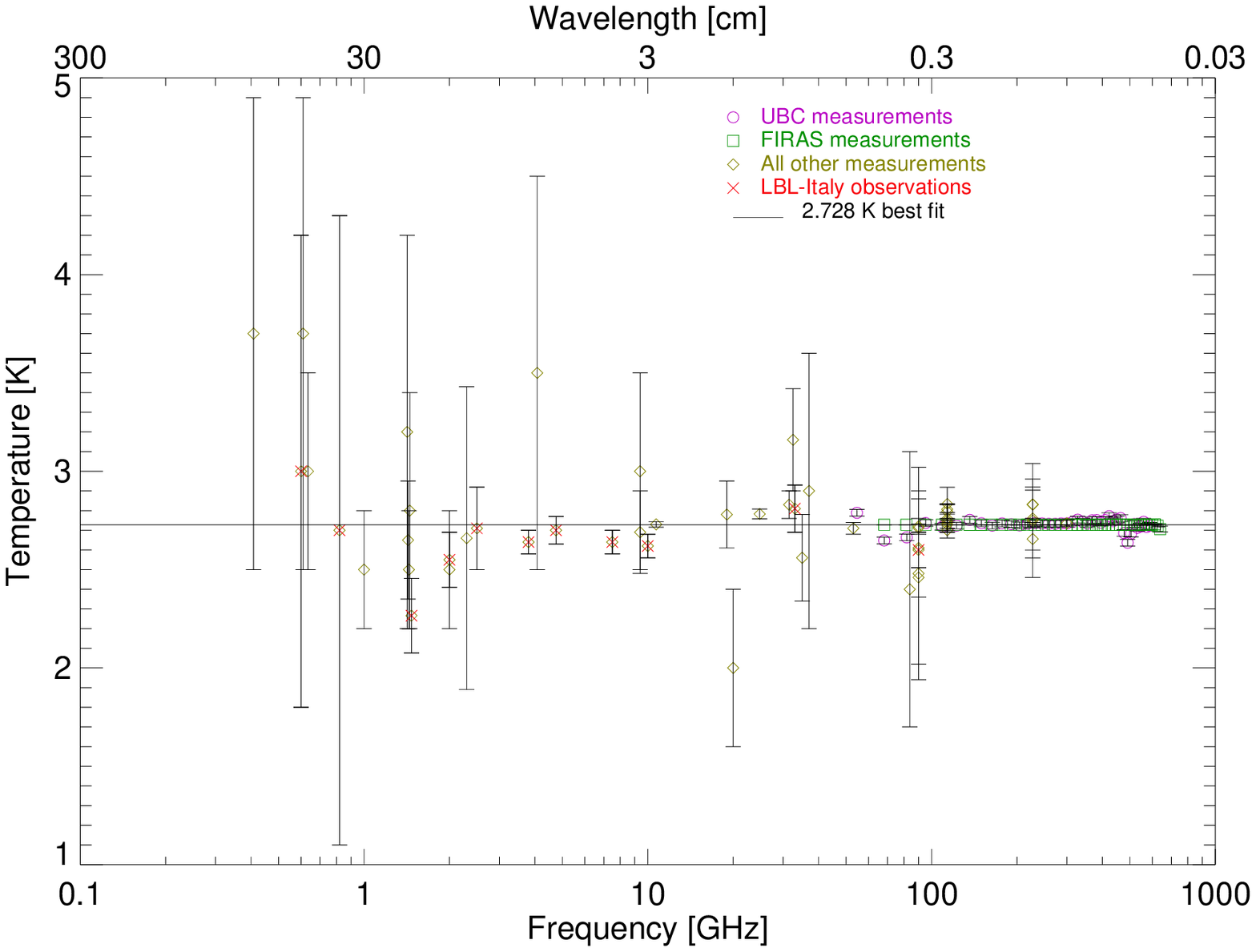}}
\noindent

\figcap{Precise measurements of the CMB spectrum, plotted in thermodynamic
temperature units (in color).  Again the line represents a 2.73~K blackbody,
and the references for the experiments can be found at the end.}

\smallskip

The resulting CMB distortion is a temperature decrement
$$
\Delta T_{\rm RJ} = -2y\;T_\gamma
 \eqno(20.2)
$$
in the Rayleigh-Jeans ($h\nu/kT \ll 1$) portion of the spectrum, 
and a rapid rise in temperature in
the Wien ($h\nu/kT \gg 1$) region, {\it i.e.}~photons are shifted from low to
high frequencies.
The magnitude of the distortion is related 
to the total energy transfer [4]
$\Delta E$ by 
$$
\Delta E/E_{\rm CBR} = e^{4y} - 1 \simeq 4y\period
 \eqno(20.3)
$$
A prime candidate for producing a Comptonized spectrum is a hot 
intergalactic medium.
A hot ($T_e>10^5\,$K) medium in clusters of galaxies can and does produce a
partially Comptonized spectrum as seen through the cluster, known as the
Sunyaev-Zel'dovich effect.
Based upon X-ray data, the predicted large angular scale total combined effect
of the hot intracluster medium should produce $y \lsim 10^{-6}$ \cite{Ceb94}.

\subsection{Bose-Einstein or chemical potential distortion} 
Early energy release ($z\sim10^5$--$10^7$).
After many Compton scatterings ($y > 1$), the photons and electrons will reach 
statistical (not thermodynamic) equilibrium, 
because Compton scattering conserves photon number.  This equilibrium is
described by the Bose-Einstein distribution with non-zero chemical
potential:
$$
n = \frac{1}{ e^{x + \mu_0} - 1 } \comma
 \eqno(20.4)
$$
where  $x \equiv { h \nu / k T}$  and
$\mu_0 \simeq 1.4\; {\Delta E} / E_{\rm CBR}$, with
$\mu_0$ being the dimensionless chemical potential that
is required.

The collisions of electrons with nuclei in the plasma produce
free-free (thermal bremsstrahlung) radiation:
$ e Z \rightarrow e   Z  \gamma $.
Free-free emission thermalizes the spectrum to the plasma temperature
at long wavelengths.
Including this effect, the chemical potential becomes frequency-dependent,
$$
\mu(x) = \mu_0 e^{- 2x_b/x }\comma
 \eqno(20.5)
$$
where $x_b$ is the transition frequency at which
Compton scattering of photons to higher frequencies
is balanced by free-free creation of new photons.
The resulting spectrum has a sharp drop in brightness temperature
at centimeter wavelengths \cite{burigana91}.
The minimum wavelength is determined by $\Omega_B$.

\vskip0.5truecm
\centerline{\epsfxsize=8.5truecm \epsfbox{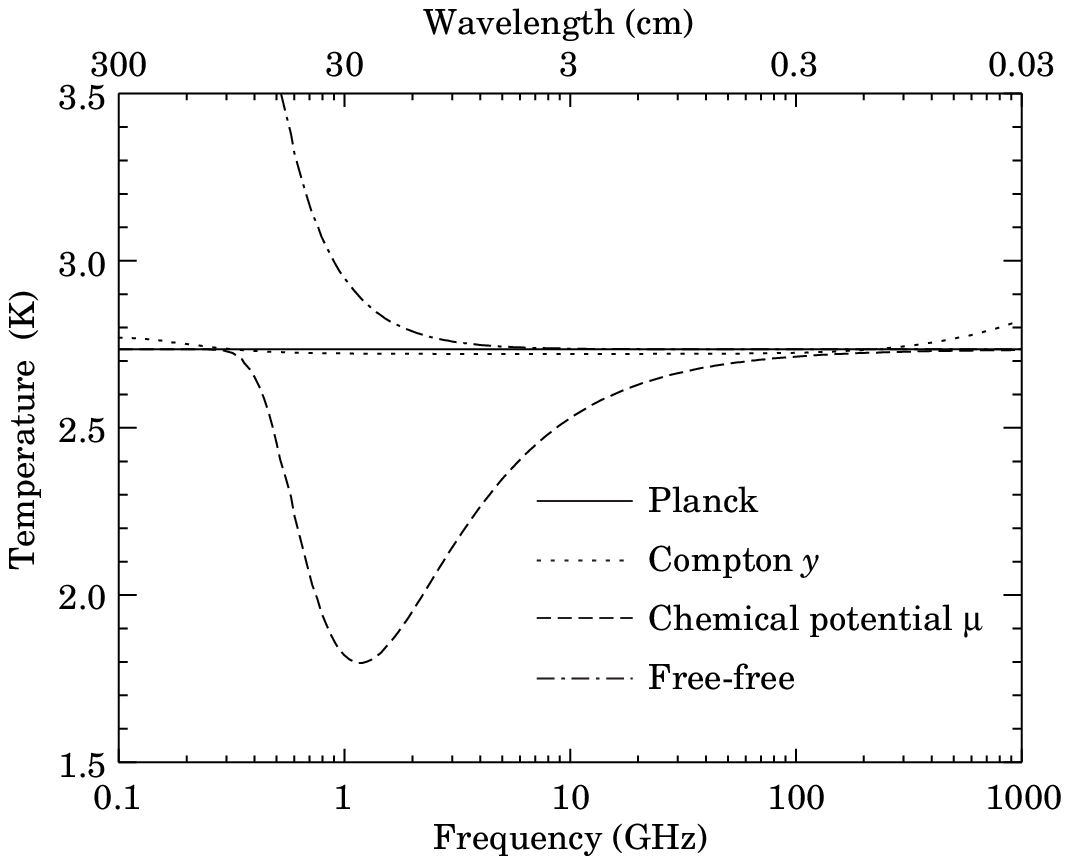}}

\noindent
\figcap{The shapes of expected, but so far unobserved, 
CMB distortions, resulting from energy-releasing processes
at different epochs.}

\smallskip

The equilibrium
Bose-Einstein distribution results from the oldest non-equilibrium
processes $(10^5 < z < 10^7)$, 
such as the decay of relic particles or primordial inhomogeneities.
Note that free-free emission (thermal bremsstrahlung) 
and radiative-Compton scattering 
effectively erase any distortions \cite{Danese82}
to a Planckian spectrum for epochs earlier than $z \sim 10^7$.

\subsection{Free-free distortion} Very late energy release ($z\ll10^3$).
Free-free emission can create rather than erase spectral distortion 
in the late universe, for recent reionization $(z < 10^3)$ and
from a warm intergalactic medium.  The distortion arises
because of the lack of Comptonization at recent epochs.
The effect on the present-day CMB spectrum is described by
$$
\Delta T_{f\!f} = T_{\gamma}\; {Y_{f\!f}}/{x^2},
 \eqno(20.6)
$$
where $T_{\gamma}$ is the undistorted photon temperature,
$x$ is the dimensionless frequency, and
$Y_{f\!f}/x^2$ is the optical depth to free-free emission:

$$
Y_{f\!f} 
= \int^z_0 ~\frac{T_e(z') - T_{\gamma}(z') }{ T_e(z') }
\frac{ 8 \pi e^6 h^2 n_e^2 \;g }{ 3 m_e (kT_{\gamma})^3\;
 \sqrt{6\pi\, m_e\, k T_e} }
\;\frac{dt}{dz'} dz'\period
 \eqno(20.7)
$$
\noindent
Here $h$ is Planck's constant,
$n_e$ is the electron density and $g$ is the Gaunt 
factor \cite{bartlett91}.

\subsection{Spectrum summary}
The CMB spectrum is consistent with a blackbody spectrum over
more than three decades of frequency around the peak.
A least-squares fit to all CMB measurements yields:

\bigskip
\vbox{
\settabs7\columns
\+&\strut$T_{\gamma} = 2.728 \pm 0.002$ K \cr
\+&\strut$n_{\gamma} = (2\zeta(3)/\pi^2) T_\gamma^3 \simeq 413\,
 {\rm cm^{-3}}$ \cr
\+&\strut$\rho_{\gamma} = (\pi^2 /15) T_\gamma^4 \simeq 4.68
 \times 10^{-34}\,
	{\rm g}\,{\rm cm^{-3}} \simeq 0.262\,{\rm eV}\,{\rm cm^{-3}}$\cr
\+&\strut$|y| < 1.2 \times 10^{-5}$ &&(95\%\ CL)\cr
\+&\strut$|\mu_0| < 9 \times 10^{-5}$ &&(95\%\ CL)\cr
\+&$\strut|Y_{f\!f}| < 1.9 \times 10^{-5}$ &&(95\%\ CL)\cr
}
\bigskip
\noindent
The limits here \cite{fixsen96} correspond to limits
[11--13]
\nocite{fixsen96}\nocite{bern94}
on energetic processes
$\Delta E / E_{\rm CBR} < ~2 \times 10^{-4}$
occurring between redshifts $10^3$ and $5 \times 10^6$
(see Fig.~20.4).  The best-fit temperature from the COBE FIRAS
experiment is $T_\gamma=2.728\pm0.002\,$K [11].

\vskip-2.75truecm
\centerline{\epsfxsize=10.75truecm \epsfbox{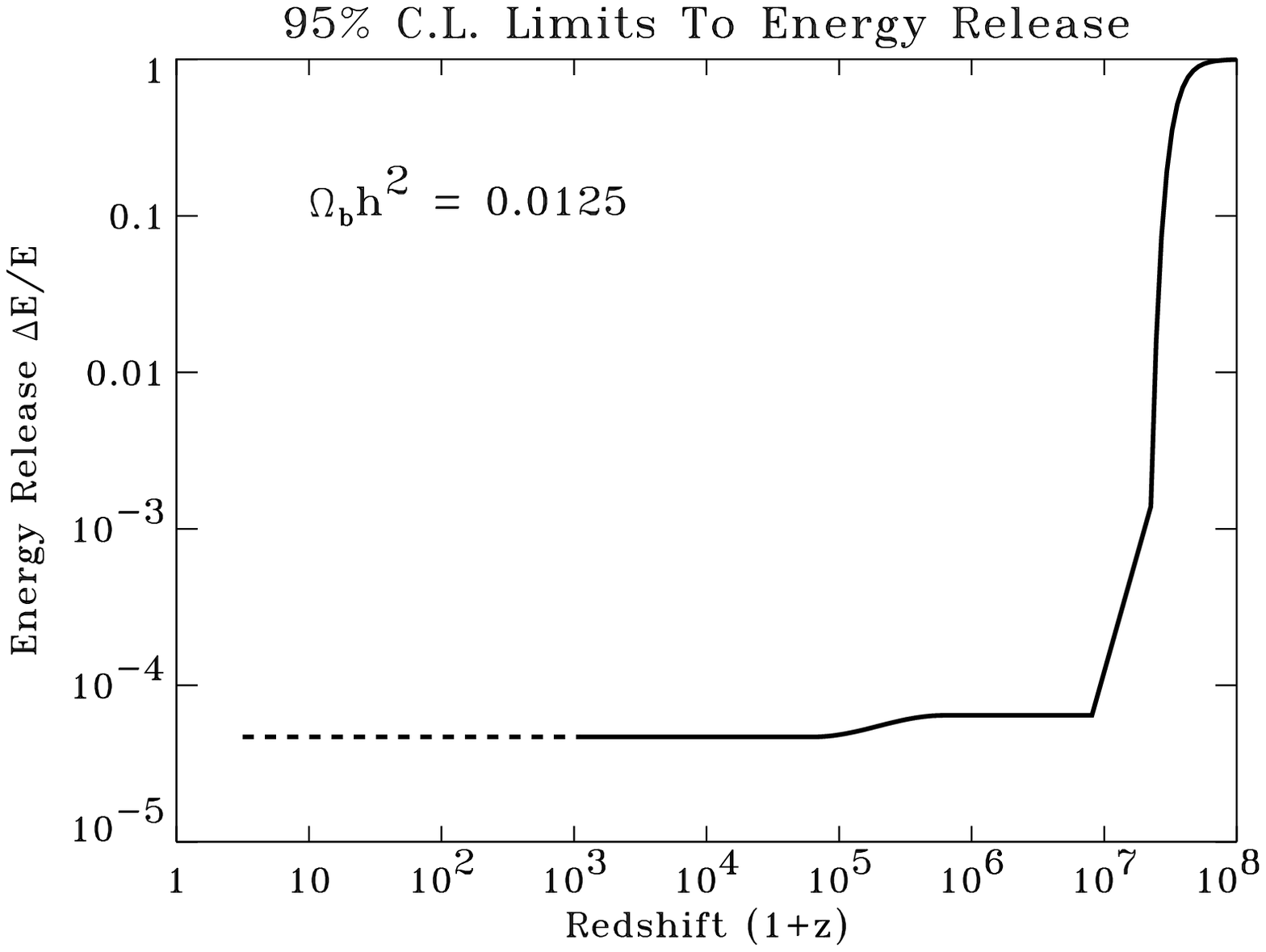}}
\vskip-0.5truecm

\noindent
\figcap{Upper Limits (95\%\ CL) on fractional energy
($\Delta E/E_{\rm CBR}$) releases 
as set by lack of CMB spectral distortions resulting from 
processes at different epochs.  These can be translated into constraints
on the mass, lifetime and photon branching ratio of unstable relic
particles, with some additional dependence on cosmological parameters
such as $\Omega_B$ \nocite{wright94}\nocite{husilk93} [9,10].}

\smallskip

\section{Deviations from isotropy}
Penzias and Wilson reported that the CMB was isotropic and unpolarized
to the 10\%\ level. Current observations show that the CMB is 
unpolarized at the $10^{-5}$ level but has a dipole anisotropy at the
$10^{-3}$ level, with smaller-scale anisotropies at the $10^{-5}$ level.
Standard theories predict anisotropies in linear polarization 
well below currently achievable
levels, but temperature anisotropies of roughly the amplitude
now being detected.

It is customary to express the CMB temperature anisotropies on the sky
in a spherical harmonic expansion,
$$
{\Delta T\over T}(\theta,\phi) = \sum_{\ell m} a_{\ell m}
Y_{\ell m}(\theta, \phi) \comma
 \eqno(20.8)
$$
and to discuss the various multipole amplitudes.  The power at a given
angular scale is roughly $\ell\sum_m\left|a_{\ell m}\right|^2/4\pi$,
with $\ell\sim1/\theta$.

\subsection{The dipole}
The largest anisotropy is in the 
$\ell=1$ (dipole) first spherical harmonic, with amplitude
at the level of $\Delta T / T = 1.23 \times 10^{-3}$.
The dipole is interpreted as the result of the Doppler shift caused
by the solar system motion relative to the nearly isotropic blackbody field.
The motion of the observer (receiver)
with velocity $\beta = v/c$ relative
to an isotropic Planckian radiation field of temperature ${T_0}$ produces
a Doppler-shifted temperature
$$
\eqalignno{
T(\theta) &= T_0 (1 - \beta^{2})^{1/2}/(1 - \beta \cos\theta)  \cr
&= T_0 \, \left(1 + \beta \cos\theta + (\beta^{2}/2) \cos2\theta
+ O(\beta^3)\right )\period
 &(20.9)
\cr}
$$

The implied velocity \cite{fixsen96} for
the solar-system barycenter
is $\beta = 0.001236 \pm 0.000002$ (68\% CL)
or $v = 371\pm 0.5\,{\rm km}\,{\rm s}^{-1}$,
assuming a value $T_0 = 2.728 \pm 0.002\,$K,
towards
$(\alpha,\delta) = (11.20^{\rm h} \pm 0.01^{\rm h}, -7.22^{\circ}
\pm 0.08^{\circ} $),
or $(\ell,b) = (264.31^{\circ}\pm0.17^{\circ}, 48.05^{\circ}\pm0.10^{\circ} $).
Such a solar-system velocity implies a
velocity for the Galaxy and the Local Group of galaxies relative
to the CMB. The derived velocity is
$v_{\rm LG} = 627 \pm 22\,{\rm km}\,{\rm s}^{-1}$ toward
$(\ell,b) 
= (276^{\circ} \pm  3^{\circ}, 30^{\circ} \pm 3^{\circ} $),
where most of the error 
comes from uncertainty in the velocity of the solar
system relative to the Local Group.

The Doppler effect of this velocity and of the velocity of the Earth
around the Sun, as well as any velocity of the receiver relative to the Earth,
is normally removed for the purposes of CMB anisotropy study.
The resulting high degree of CMB isotropy is the strongest evidence 
for the validity of the Robertson-Walker metric.

\vskip-1.5truecm
\centerline{\epsfxsize=8.75cm \epsfbox{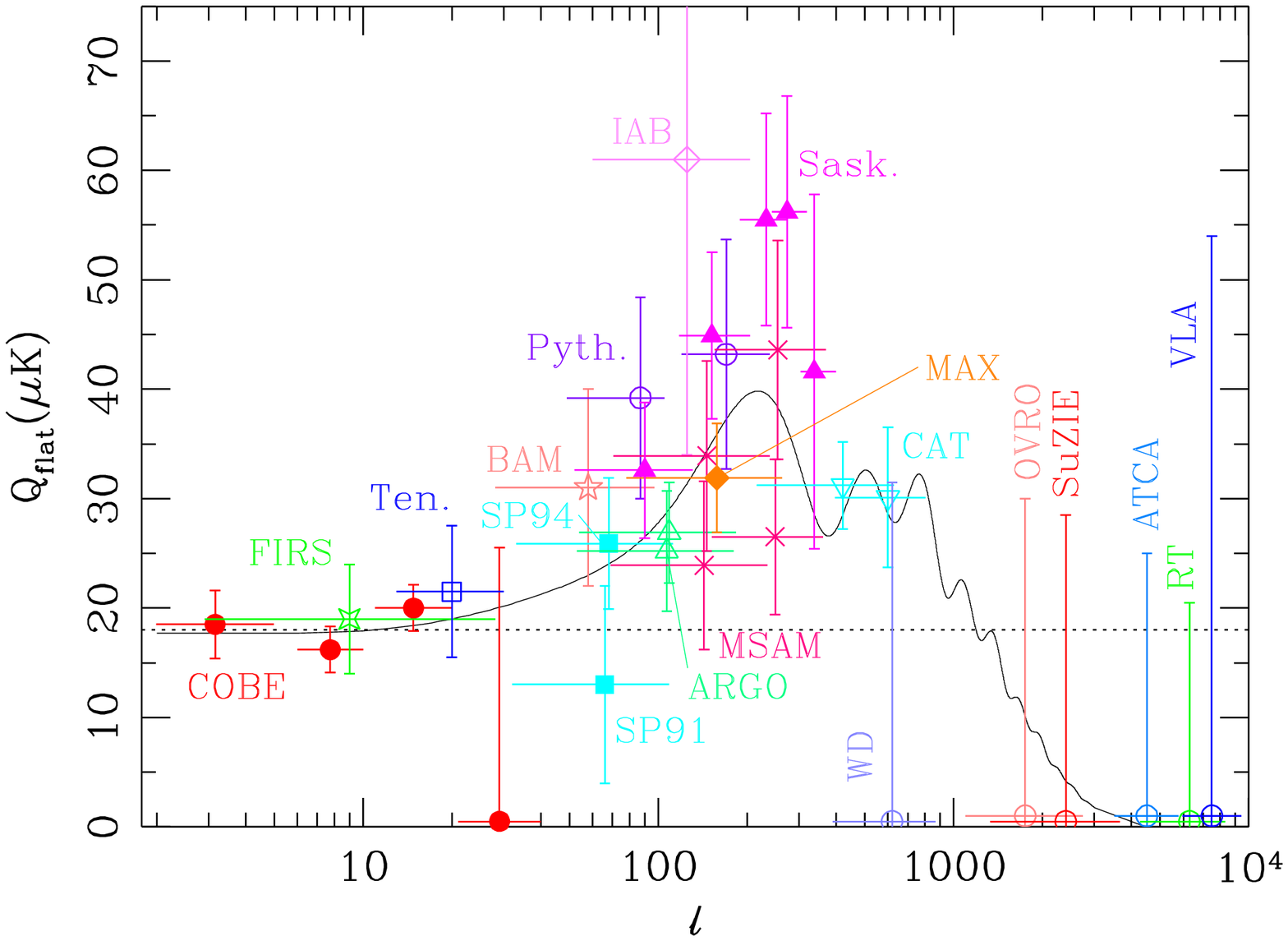}}

\noindent
\figcap{Current status of CMB anisotropy observations, 
adapted from Scott, Silk, \&  White (1995) \cite{scott95}.
This is a (color!)
representation of the results from COBE, together with a wide
range of ground- and balloon-based experiments which have operated in the
last few years.
Plotted are the quadrupole amplitudes
for a flat (unprocessed scale-invariant spectrum of primordial perturbations,
\ie, a horizontal line) anisotropy spectrum
that would give the observed results for each experiment.
In other words each point is the normalization of a flat spectrum derived
from the individual experiments.
The vertical error bars represent estimates of 68\% CL, while
the upper limits are at 95\% CL.  Horizontal bars
indicate the range of $\ell$ values sampled.
The curve indicates the expected spectrum for a standard CDM model
($\Omega_0 = 1, \Omega_B =0.05, h =0.5$), although true comparison with
models should involve convolution of this curve with each experimental filter
function.  (References for this figure are at the end of this section under
``CMB Anisotropy References.'')}

\smallskip

\subsection{The quadrupole}
The rms quadrupole anisotropy amplitude is defined through
$Q_{\rm rms}^2/T_\gamma^2=\sum_m\left|a_{2 m}\right|^2/4\pi$.
The current estimate of its value is
$4\,\mu{\rm K}\le Q_{\rm rms}\le 28\,\mu{\rm K}$ for a 95\% confidence
interval \cite{bennett96}.
The uncertainty here includes both statistical
errors and systematic errors, which are dominated by 
the effects of galactic emission modelling.
This level of quadrupole anisotropy allows one to set general
limits on anisotropic expansion, shear, and vorticity; all such
dimensionless quantities are constrained to be less than about
$10^{-5}$.
For specific homogeneous cosmologies, fits to the whole anisotropy pattern
allow stringent limits to be placed on, for example, the global rotation at
the level of about $10^{-7}$ of the expansion rate 
\cite{KogHinBan97}.

\smallskip
\subsection{Smaller angular scales}
The COBE-discovered \cite{smoot92}
higher-order ($\ell > 2$) anisotropy is interpreted 
as being the result
of perturbations in the energy density of the early Universe,
manifesting themselves at the epoch of the CMB's last scattering.
Hence the detection of these anisotropies has provided evidence for
the existence of primordial density perturbations which grew through
gravitational instability to form all the structure we observe today.

\vskip-1.5truecm
\centerline{\epsfxsize=8.75cm \epsfbox{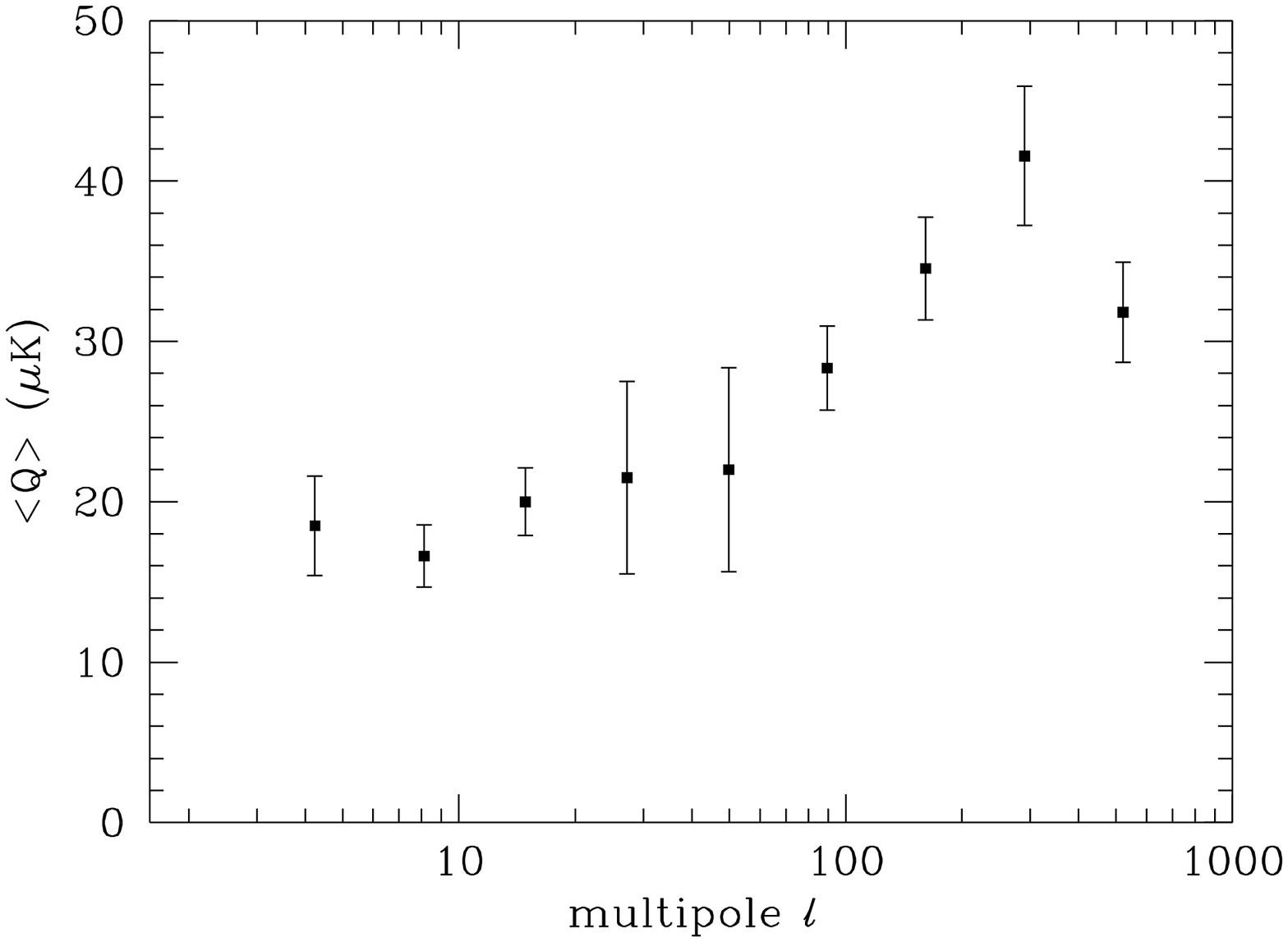}}

\noindent
\figcap{This is a binned version of the previous figure.  To obtain this
figure we took all reported detections, split the multipole range into
equal logarithmic `bins', and calculated the weighted average in each bin.
Although this is not a statistically rigorous procedure, the resulting
figure gives a visual indication of the current consensus.  It is
also worth mentioning that there is no strong indication for
excess scatter (above Gaussian) within each bin.}

\smallskip

In the standard scenario the last scattering takes place at a redshift
of approximately 1100, at which epoch the large
number of photons was no longer able to keep the hydrogen sufficiently ionized.
The optical thickness of the cosmic photosphere is roughly $\Delta z \sim 100$
or about 5 arcminutes, 
so that features smaller than this size are damped.

Anisotropies are observed on angular scales larger than this 
damping scale (see Fig.~20.5 and 20.6),
and are consistent with those expected from an initially scale-invariant
power spectrum (flat = independent of scale) of potential and 
thus metric fluctuations.
It is believed that the large scale structure in the Universe
developed through the process of gravitational instability,
where small primordial perturbations in energy density were amplified
by gravity over the course of time.
The initial spectrum of density perturbations
can evolve significantly in the epoch $z > 1100$ for causally connected
regions (angles $\lsim 1^\circ\;\Omega_{\rm tot}^{1/2}$).
The primary mode of evolution
is through adiabatic (acoustic) oscillations, leading to a series 
of peaks that encode information about
the perturbations and geometry of the Universe, as well as
information on $\Omega_0$, $\Omega_B$, $\Omega_\Lambda$ (cosmological
constant), 
and $H_0$ [18].
The location of the first acoustic peak
is predicted to be at $\ell \sim 220 \;\Omega_{\rm tot}^{-1/2}$ or 
$\theta \sim 0.3^\circ \;\Omega_{\rm tot}^{1/2}$
and its amplitude increases is a calculable function of the parameters.

Theoretical models generally predict a power spectrum 
in spherical harmonic amplitudes,
since the models lead to primordial fluctuations 
and thus $a_{\ell m}$ that are Gaussian random fields,
and hence
the power spectrum in  $\ell$ is sufficient to characterize the results.
The power at each $\ell$ is $(2 \ell +1) C_\ell/(4\pi)$,
where $C_\ell \equiv \left\langle{|a_{\ell m}|^2}\right\rangle$
and a statistically isotropic sky means that all $m$s are equivalent.
For an idealized full-sky observation, the variance of 
each measured $C_\ell$ is $[2 /(2 \ell +1 )] C^2_\ell$.
This sampling variance (known as cosmic variance) comes about because
each $C_\ell$ is chi-squared distributed with $( 2 \ell +1 )$ degrees
of freedom for our observable volume of the Universe \cite{white94}.
Thomson scattering of the anisotropic radiation field also generates linear
polarization at the roughly 5\% level \cite{HuWhite97}.
Although difficult to detect, the
polarization signal should act as a strong confirmation of the general
paradigm.

Fig.~20.7 shows the theoretically predicted anisotropy
power spectrum
for a sample of models,
plotted as $\ell(\ell+1) C_\ell$ versus $\ell$ which is the power per
logarithmic interval in $\ell$ or, equivalently, the two-dimensional 
power spectrum.
If the initial power spectrum of perturbations is the result of 
quantum mechanical fluctuations produced and 
amplified during inflation, then the shape of the anisotropy spectrum 
is coupled to the ratio of contributions from 
density (scalar) and gravity wave (tensor) perturbations \cite{Lidsey}.
If the energy scale of inflation at the appropriate epoch is at
the level of $\simeq 10^{16}$GeV, then detection of
the effect of gravitons is possible,
as well as partial reconstruction of the inflaton potential.
If the energy scale is $\lsim 10^{14}$GeV, then density
fluctuations dominate and less constraint is possible.

\vskip-1.5truecm
\centerline{\epsfxsize=8.75cm \epsfbox{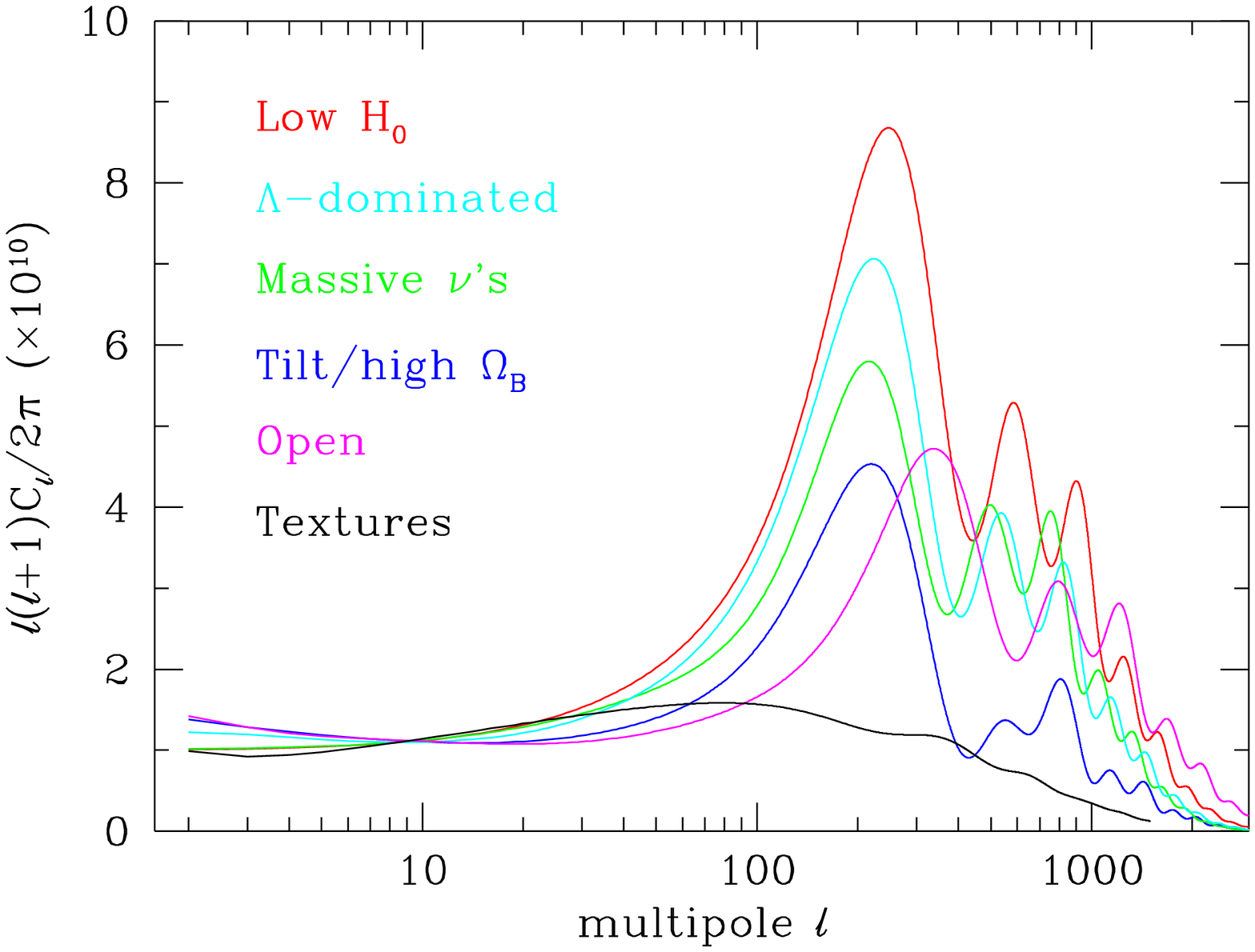}}
\noindent

\figcap{Examples of theoretically predicted $\ell(\ell+1)C_\ell$ 
or CMB anisotropy power spectra \cite{SelZar96}.
The plot indicates that precise measurements
of the CMB anisotropy power spectrum could distinguish between
models which are currently favored from galaxy clustering and other
considerations.  The textures model is from Ref.\cite{ZalSpeSel97}.}
\smallskip
 
Fits to data over smaller angular scales are often quoted as the expected
value of the quadrupole $\left\langle Q\right\rangle$
for some specific theory, \eg~a model with power-law
initial conditions (primordial 
density perturbation power spectrum $P(k)\propto k^n$).
The full 4-year COBE DMR data give
$\left\langle Q\right\rangle=15.3^{+3.7}_{-2.8}\;\mu$K, after projecting out
the slope dependence, while the best-fit slope is
$n=1.2\pm0.3$, and for a pure  $n=1$ 
(scale-invariant potential perturbation) spectrum
$\left\langle Q\right\rangle
(n=1)=18\pm1.6\,\mu$K [15,24]\nocite{gorski96}.
The conventional notation
is such that $\left\langle Q\right\rangle^2/T_\gamma^2=5C_2/4\pi$,
and an alternative convention is to plot the ``band-power''
$\sqrt{\ell(2\ell+1)C_\ell/4\pi})$.
The fluctuations measured by other experiments can also be quoted in terms
of $Q_{\rm flat}$, the equivalent value of the quadrupole for a flat ($n=1$)
spectrum, as presented in Fig.~20.5.

It now seems clear that there is more power at sub-degree scales than at COBE
scales (see Fig.~20.5),
which provides some model-dependent information on cosmological
parameters [18,25]\nocite{KogHin96}, for example $\Omega_B$.
In terms of such parameters,
fits to the COBE data alone yield $\Omega_0>0.34$ at 95\% CL \cite{YamBun96}
and $\Omega_{\rm tot}<1.5$ also at 95\% CL \cite{WhiSco96}, for inflationary
models.
Only somewhat weak conclusions can be drawn based on the current
smaller angular scale data (see Fig.~20.5).  
A sample preliminary fit \cite{LinBar97} finds 
$\Omega_0h^{1/2}\simeq0.55\pm0.10$ (68\% CL).

However, new data are being acquired at an increasing rate, 
with a large number of improved ground- and balloon-based experiments 
being developed.  
It appears that we are not far from being able
to distinguish crudely between currently favored models, 
and to begin a more precise determination of cosmological parameters.
A vigorous suborbital and interferometric 
program could map out the CMB anisotropy power spectrum
to about 10\%\ accuracy and determine several parameters
at the 10 to 20\%\ level in the next few years.

There are also now two approved satellite missions: the NASA Microwave
Anisotropy Probe (MAP), scheduled for launch in 2000; snd the ESA
Planck Surveyor, expected for around 2004.  The improved sensitivity, freedom
from earth-based systematics and all-sky coverage allow a simultaneous
determination of many of the cosmological parameters to unprecedented
precision: for example, $\Omega_0$ and $n$ to about 1\%, $\Omega_B$ and
$H_0$ at the level of a few per cent \cite{Jungman96}.

Furthermore, detailed measurement of the polarization signal provides more
precise information on the physical parameters.  In particular it allows a
clear distinction of any gravity wave contribution, which is crucial to probing
the $\sim10^{16}$GeV energy range.
The fulfillment of this promise may await an
even more sensitive generation of satellites.

\medskip
\medskip
\noindent{\bf Acknowledgements}

We would like to thank the numerous colleagues who helped in compiling and
updating this review, in particular those who sent us
details of experimental results ahead of publication.  We are also greatful
for detailed comments from Martin White and the referees, and for
the diligence of the RPP staff, particularly Don Groom and Betty
Armstrong.  This work was partially supported by the US DOE and the
Canadian NSERC.

\medskip
\medskip
\noindent{\bf References:}

\parindent=0pt
\parskip=0pt plus 1pt minus 1pt

\pp
1. R.A. Alpher and R.C. Herman, Physics Today, Vol.\ 41, No.\ 8, p.~24 (1988)

\pp
2. A.A. Penzias and R. Wilson, \aj142,419 (1965);

\ppextra
R.H. Dicke, P.J.E. Peebles, P.G. Roll, and D.T. Wilkinson,
\aj142,414 (1965)

\pp
3. P.J.E. Peebles, ``Principles
of Physical Cosmology,'' Princeton U.\ Press, p.~168 (1993)

\pp
4. R.A. Sunyaev and Ya.B. Zel'dovich, \araa18,537 (1980)

\pp
5. M.T. Ceballos and X. Barcons, MNRAS {\bf 271}, 817 (1994)

\pp
6. C. Burigana, L. Danese, and G.F. De Zotti, \aap246,49 (1991) 

\pp
7. L. Danese and G.F. De Zotti,  \aap107,39 (1982);

\ppextra
G. De Zotti, \ppnp17,117 (1987)

\pp
8. J.G. Bartlett and A. Stebbins, \aj371,8 (1991)

\pp
9. E.L. Wright \etal, \aj420,450 (1994)

\pp
10. W. Hu and J. Silk, \prl70,2661 (1993)

\pp
11. D.J. Fixsen \etal, \aj473,576 (1996)

\pp
12. J.C. Mather \etal, \aj420,439 (1994)

\pp
13. M. Bersanelli \etal, \aj424,517 (1994)

\pp
14. A. Kogut \etal, \aj419,1 (1993);

\ppextra
C. Lineweaver \etal, \aj470,L38 (1996)

\pp
15. C.L. Bennett \etal, \aj464,L1 (1996)

\pp
16. E.F. Bunn, P. Ferreira, and J. Silk, \prl77,2883 (1996);
\ppextra
A. Kogut, G. Hinshaw, and A.J. Banday, Phys. Rev. {\bf D55}, 1901 (1997)

\pp
17. G.F. Smoot \etal,  \aj396,L1 (1992)

\pp
18. D. Scott, J. Silk, and M. White, \sci268,829 (1995);

\pp
W. Hu, J. Silk, and N. Sugiyama, \nat386,37 (1996)

\pp
19. M. White, D. Scott, and J. Silk, \araa32,329 (1994)

\pp
20. W. Hu and M. White, New Astron. {\bf 2}, 323 (1997)

\pp
21. J.E. Lidsey \etal, Rev. Mod. Phys. {\bf 69}, 373 (1997);
\ppextra
D.H. Lyth, Phys. Rep., in press (1997) (hep-ph/9609431)

\pp
22. U. Seljak, M. Zaldarriaga, \aj469, 437 (1996)

\pp
23. U.-L. Pen, U. Seljak, and N. Turok, \prl79,1611 (1997)

\pp
24. K.M. G{\'o}rski \etal, Astrophys.\ J., 464, L11 (1996)

\pp
25. A. Kogut and G. Hinshaw, Astrophys.\ J.\ Lett., 464, L39 (1996)

\pp
26. K. Yamamoto and E.F. Bunn, Astrophys.\ J., 464, 8 (1996)

\pp
27. M. White and D. Scott, \aj459,415 (1996)

\pp
28. C.H. Lineweaver and D. Barbosa, Astrophys.\ J., submitted
(astro-ph/9706077)

\pp
29. G. Jungman, M. Kamionkowski, A. Kosowsky, and D.N. Spergel,  Phys.\ Rev.\
{\bf D54}, 1332 (1996);
\ppextra
W. Hu and M. White, Phys.\ Rev.\ Lett., 77, 1687 (1996);
\ppextra
J.R. Bond, G. Efstathiou, and M. Tegmark, MNRAS, in press (1997)
(astro-ph/9702100);
\ppextra
M. Zaldarriaga, D. Spergel, and U. Seljak, \aj488,1 (1997)
\medskip

\parindent=0pt
\noindent{\bf CMB Spectrum References:}
\medskip

\pp
{\bf FIRAS:\rm} J.C. Mather \etal, \aj420,439 (1994);
\ppextra
D. Fixsen \etal, \aj420,445 (1994);
\ppextra
D. Fixsen \etal, Astrophys.\ J., in press (1996)

\pp
{\bf DMR:\rm} A. Kogut \etal, \aj419,1 (1993);
\ppextra
A. Kogut \etal, Astrophys.\ J., submitted (1996)

\pp
{\bf UBC:\rm} H.P. Gush, M. Halpern, and E.H. Wishnow, \prl65,537 (1990)

\pp
{\bf LBL-Italy:\rm} G.F. Smoot \etal, \prl51,1099 (1983);
\ppextra
M. Bensadoun \etal, \aj409,1 (1993);
\ppextra
M. Bersanelli \etal, \aj424,517 (1994);
\ppextra
M. Bersanelli \etal, Astrophys.\  Lett.\  and Comm.\  {\bf 32}, 7 (1995);
\ppextra
G. De~Amici \etal, \aj381,341 (1991);
\ppextra
A. Kogut \etal, \aj335,102 (1990);
\ppextra
N. Mandolesi \etal, \aj310,561 (1986);
\ppextra
G. Sironi, G. Bonelli, and M. Limon, \aj378,550 (1991)

\pp
{\bf Princeton:\rm} S. Staggs \etal, 
Astrophys.\ Lett.\ \& Comm.\ {\bf 32}, 3 (1995);
\ppextra
D.G. Johnson and D.T. Wilkinson, \aj313,L1 (1987)

\pp
{\bf Cyanogen:\rm} K.C. Roth, D.M. Meyer, and I. Hawkins, \aj413,L67 (1993);
\ppextra
K.C. Roth and  D.M. Meyer, \aj441,129 (1995);
\ppextra
E. Palazzi \etal, \aj357,14 (1990)
\medskip

\parindent=0pt
\noindent{\bf CMB Anisotropy References:}
\medskip

\pp
{\bf COBE:\rm}
G. Hinshaw \etal, \aj464,L17 (1996)

\pp
{\bf FIRS:\rm} K. Ganga, L. Page, E. Cheng, and S. Meyers,  \aj432,L15 (1993)

\pp
{\bf Ten.:\rm} C.M.Guti{\'e}rrez \etal,\aj480,L83 (1997)

\pp
{\bf BAM:\rm} G.S. Tucker \etal, \aj475,L73 (1997)

\pp
{\bf SP91:\rm} J. Schuster \etal, \aj412,L47 (1993). (Revised, see
{\bf SP94\rm} reference.)

\pp
{\bf SP94:\rm} J.O. Gundersen \etal, \aj443,L57 (1994)

\pp
{\bf Sask.:\rm} C.B. Netterfield \etal, \aj474,47 (1997)

\pp
{\bf Pyth.:\rm} S.R. Platt \etal, \aj475,L1 (1997)

\pp
{\bf ARGO:\rm} P. de Bernardis \etal, \aj422,L33 (1994);
\ppextra
S. Masi \etal,\aj463,L47 (1996)

\pp
{\bf IAB:\rm} L. Piccirillo and P. Calisse,  \aj413,529 (1993)

\pp
{\bf MAX:\rm} S.T. Tanaka \etal, \aj468,L81 (1996);
M. Lim \etal, \aj469,L69 (1996)

\pp
{\bf MSAM:\rm} E.S. Cheng \etal, \aj456,L71 (1996);
E.S. Cheng \etal, Astrophys. J., in press (1997)
(astro-ph/9705041)

\pp
{\bf CAT:\rm} P.F.S. Scott \etal, \aj461,L1 (1996)
J.C. Baker, Proc. {\it Particle Physics and Early Universe
Conference},
(1997) (http://www.mrao.cam.ac.uk/ppeuc/astronomy/papers/baker
/baker.html);
\ppextra
J.C. Baker et al., in preparation

\pp
{\bf WD:\rm} G.S. Tucker, G.S. Griffin, H.T. Nguyen, and J.B. Peterson,
      \aj419,L45 (1993)

\pp
{\bf OVRO:\rm} A.C.S. Readhead \etal, \aj346,566 (1989)

\pp
{\bf SuZIE:\rm} S. E. Church \etal, \aj484,523 (1997)

\pp
{\bf ATCA:\rm} R. Subrahmayan, R.D. Ekers, M. Sinclair, and J. Silk, 
MNRAS {\bf 263}, 416 (1993)

\pp
{\bf RT:\rm} M.E. Jones, Proc. {\it Particle Physics and Early Universe
Conference} (1997)
(http://www.mrao.cam.ac.uk/ppeuc/
astronomy/papers/jones/jones.html);
\ppextra
M.E. Jones et al., in preparation

\pp
{\bf VLA:\rm} B. Partridge \etal, \aj483,38 (1997)
  
\bye